\documentclass[twocolumn,preprintnumbers,amsmath,amssymb,aps,prb,longbibliography]{revtex4-1}
\usepackage{graphicx}
\usepackage{color}
\usepackage{dcolumn}
\usepackage{bm}
\usepackage{hyperref}
\begin{document}

\title{Characterizing Different Motility Induced Regimes in Active Matter with Machine Learning and Noise}  
\author{
 D. McDermott$^{1}$, 
 C. Reichhardt$^{2}$,
 and C. J. O. Reichhardt$^{2}$
} 
\affiliation{
$^1$X-Theoretical Design Division, Los Alamos National Laboratory, Los Alamos, New Mexico 87545 USA\\ 
$^2$Theoretical Division,  
Los Alamos National Laboratory, Los Alamos, New Mexico 87545 USA\\ 
}

\date{\today}
\begin{abstract}
  We examine motility-induced phase separation (MIPS) in two-dimensional run and tumble disk systems using both machine learning and noise fluctuation analysis. Our measures suggest that within the MIPS state there are several distinct regimes as a function of density and run time, so that systems with MIPS transitions exhibit an active fluid, an active crystal, and a critical regime. The different regimes can be detected by combining 	an order parameter extracted from principal component analysis with a cluster stability measurement. The principal component-derived order parameter is maximized in the critical regime, remains low in the active fluid, and has an intermediate value in the active crystal regime. We demonstrate that machine learning can better capture dynamical properties of the MIPS regimes compared to more standard structural measures such as the maximum cluster size. The different regimes can also be characterized via changes in the noise power of the fluctuations in the average speed. In the critical regime, the noise power passes through a maximum and has a broad spectrum with a $1/f^{1.6}$ signature, similar to the noise observed near depinning transitions or for solids undergoing plastic deformation.
\end{abstract}
\maketitle
    
\section{Introduction}
    
Active matter denotes
systems composed of self-propelling agents or particles
that move using
internal driving or energy harvested from the surrounding environment
\cite{Ramaswamy10,Marchetti13,Bechinger16}.
Examples of active matter include
bacteria~\cite{Sokolov12},
engineered systems such as
robots~\cite{Scholz18}, and colloidal particles that move
using a variety of mechanisms \cite{Palacci13,Wang15,Bechinger16} such as
light~\cite{Buttinoni13,Schmidt19} or magnets~\cite{Demirors18}.  
One of the most studied phenomena found in models
of interacting active particles, such as active disks,
is motility-induced phase segregation (MIPS),
which occurs
even for systems with only repulsive interactions
when the persistence of the motion is
large enough. In MIPS, 
for densities well below those at which the system can uniformly crystallize,
the system phase-separates into 
a dense or crystalline phase coexisting with a low density
gas \cite{Fily12}. 
Although MIPS has generally been regarded
as a single phase~\cite{Redner13,Palacci13,Buttinoni13,Cates15},
its intrinsically dynamic nature means that there could
be different dynamic regimes
or changing structures within the MIPS state.
    
Various studies
have noted interruptions to MIPS
due to mechanisms such as polar alignment
between neighboring particles~\cite{vanderLinden19},
inertia~\cite{Dai20},
or large-scale shear motions due to the high
speed of individual particles \cite{Kichatov21}.
Work on low-density, apolar active matter has shown
the cluster size and average speed of active colloids
exhibit power law distributions \cite{Ginot18}.
Experiments have revealed
a variety of phases in suspensions of
bacteria \cite{Sokolov12}
and run-and-tumble Quincke rollers \cite{Karani19},
though in these systems 
the hydrodynamics of the suspending fluids may play a role.
Furthermore, torque can mediate an
active clustered phase distinct from MIPS ~\cite{Zhang21}.
In very dense active matter, 
different kinds of intermittency have
been found as a function of activity
when the system is a solid
\cite{Mandal20}.

To characterize MIPS, 
a number of traditional measurements are commonly used
that have been adapted from equilibrium systems,
such as average cluster size,
the radial distribution function,
or the amount of bond orientational order~\cite{Steinhardt83}.
Such measures were designed to
distinguish a variety of transitions
including short versus long-range order
or order-order structural phase transitions; however,
these measures may fail to detect notable features due to dynamic changes
or the fluctuating inhomogeneity of particle environments.
As a result, the measures
may give the same signal for states where distinct types of dynamical
motion is occurring if the static structures in these
states are sufficiently similar.

Machine learning (ML) is being employed to an increasing extent as
a method for
characterizing active matter
\cite{Martiniani17,Cichos20}.
In some ML approaches, a variety of factors are combined
into a single optimized order parameter, $P_1$, 
that can
make it possible to
identify different dynamics in strongly nonequilibrium systems.
In previous work, machine learning has been used to
discern phases in soft non-active granular disks
and a variety of equilibrium \cite{Jadrich18}
and nonequilibrium systems \cite{Jadrich18a},
as well as to find transitions between different
dynamical nonequilibrium phases in driven systems \cite{McDermott20}.

Another frequently used characterization method
for nonequilibrium driven systems is
measuring features in noise fluctuations
\cite{Weissman88,Sethna01},
such as the Barkhausen noise in
driven magnetic systems
\cite{Durin05}.
Similar measures have been applied to sliding charge density waves
\cite{Gruner88},
the depinning of superconducting vortices
\cite{Marley95,Field95,Olson98a,Okuma08}, driven systems
with quenched disorder \cite{Reichhardt17},
and glass transitions in electron systems
\cite{Jaroszynski02,Reichhardt04}.
The power spectrum constructed from a time series of some fluctuating
quantity
can show transitions
from a narrow band
signature, indicative of
the presence of a characteristic frequency,
to broad band or $1/f^\alpha$ noise characteristics,
where the exponent $\alpha$ can
range from $\alpha=0$, associated with white noise,
to values as large as $\alpha=2.0$ in many systems.
The noise power, $S_0$, 
obtained by integrating the power spectrum over a particular frequency window
\cite{Rabin98},
often peaks when the system goes through an equilibrium
\cite{Chen07} or
nonequilibrium \cite{Olson98a,Okuma08} phase transition.
In the case of crackling noise,
there can be a disorder-induced critical point
at which the system exhibits a broad distribution of
avalanche sizes as well as power law
fluctuation spectra
\cite{Kuntz00,Travesset02},
and in some cases, noise measurements can be used to
identify the universality class of the underlying phase transition.
Despite the widespread use of
noise measures to characterize collective dynamics in condensed matter systems,
noise analysis has not been applied systematically to active matter.
Much of the microscopic dynamics
in active matter systems can be
accessed directly in both simulations and experiments, so
time series
of a variety of quantities should readily be available
in many of these systems.

Here we perform
numerical simulations of two-dimensional systems
of run-and-tumble active disks
at varied densities and persistence lengths both inside and
outside of the MIPS state, and show that
unsupervised machine learning and noise
analysis can be used to identify distinct dynamical regimes within
MIPS.
We show that ML is better able to
characterize dynamical behaviors
in the MIPS state than traditional
clustering measures.
Based on our ML and noise analysis,
we propose that there
are three regimes in systems that exhibit MIPS behavior.
In the active fluid or non-MIPS regime
where there is no clustering, the ML principal component $P_1$ and the
noise power $S_0$ are both low.
In the active crystal regime found for high persistence lengths
and high densities, the system has large-scale crystalline
order and both $P_1$ and $S_0$ have intermediate values.
Finally, in the critical regime
at intermediate persistence lengths and intermediate densities, both
$P_1$ and $S_0$ are large.
The principal component-derived order parameter $P_1$  
passes through its maximum value in the critical regime, and
at the same time we observe the largest
velocity noise power $S_0$ along with
broadband spectra of the form $1/f^{1.62}$.   
Both $P_1$ and $S_0$ have nonmonotonic signatures and, after
reaching maximum values, decrease as either the density or the
persistence length is increased.
For intermediate densities, the cluster stability measure also
peaks in the same persistence length
region as the peaks in $P_1$ and $S_0$ before decreasing
again at higher persistence lengths.
The  dynamics in the critical regime
resemble those of
systems exhibiting plastic depinning,
which also show a peak in the noise power and similar
$1/f^{\alpha}$ noise signatures.
At higher densities, the values of both the noise power
and the first principal component-derived order parameter are reduced.

\section{Simulation and System}
\label{sec:simulation}
We model the active particles as $N_d$ disks
interacting via excluded volume with no polar alignment.
The disks
are confined in a
two-dimensional (2D) system of size
$L \times L$ with side length $L=60.0$ and with periodic
boundary conditions in the $x$ and $y$ directions.
Each disk has a fixed radius of $r_d=0.5$.
The disk positions are initialized at
randomly chosen non-overlapping locations, and the system is
allowed to evolve until it reaches
steady state behavior.
We vary the packing fraction or area density of the disks
$\phi = \pi r_d^2 N_d / L^2$ 
by changing N$_d$ 
while holding the disk radius and sample size fixed.
For most of the results presented here, we work at
densities 
$\phi$ that are well below the jamming density 
$\phi_J \approx 0.9$
of monodisperse passive disks~\cite{Reichhardt14}.

Each disk self-propels under an active force $\vec{F}_{i}^{a}$
and runs in a randomly chosen constant direction ${\bf{\hat m}}_i$
for $\tau$ simulation time steps, where $\tau$ is the
persistence time,
before performing an instantaneous tumble and selecting a
new randomly chosen direction.
We fix the magnitude of the active force to $F^a = 1.0$ for each disk
and
initialize the clock of each particle, which tracks the amount of time
since the last tumble, to random values in the range $[0,\tau]$ so
that individual disks tumble at 
randomly distributed times.
In the absence of other interactions,
a disk would travel a distance $r_l=F^a \tau \Delta t$
during a single run interval,
where the simulation time step is $\Delta t = 0.002$.

Disk-disk interactions are modeled
by a short range harmonic repulsive force
\begin{equation}
  \vec{F}_{ij}^{dd} = k_d (\vec{r}_i - \vec{r}_j) \Theta(|\vec{r}_i-\vec{r_j}|-2r_d)  
\end{equation}
for disks with positions $\vec{r}_i$ and $\vec{r}_j$,
where the
spring constant $k_d = 50$ 
and
$\Theta$ is the Heaviside step function.
The overdamped equation of motion for an individual disk $i$ is
\begin{equation}
  \eta \vec{v}_{i} = \vec{F}_{i}^{a} + \sum_{i\neq j}^{N_{d}} \vec{F}_{ij}.
\end{equation}
where $\eta=1$ is the damping constant
representing the viscosity of the implicit suspending liquid.
We integrate the equations of motion
with a standard Verlet algorithm.

\section{Principal Component Analysis of Disk Systems}
\label{sec:PCA} 
For the machine learning, we employ
principal component analysis (PCA)~\cite{Abdi10},
which calculates 
directions of maximum variance in a data set
comprised of $M$ samples of $N$ measurements. 
PCA can identify patterns in large data sets
and perform dimensionality reduction tasks.
In condensed matter applications,
it has been shown that
an order parameter derived from PCA 
can be used to identify phase transitions
in on-lattice, Ising spin type systems 
by operating directly with matrices describing
spin states ~\cite{Carrasquilla17,Hu17}. 
To perform PCA on off-lattice systems 
of interacting disks, 
we construct matrices containing 
the instantaneous
interparticle distances, 
$r_{ij} = |{\vec r}_{i}-{\vec r}_j|$.
The measurement is described in detail in Refs.~\cite{Jadrich18,McDermott20}.
The distance matrices are calculated for a subset of probe
particles rather than for all particles in the system, and
it is important to
sort the interparticle distances in
order to construct the feature vector $\vec{f}_i$
for each probe particle $i$:
\begin{equation}
  \vec{f}_i = [r_{i0},r_{i1},r_{i2}, ..., r_{ij}, ..., r_{in}].
 \label{eq:feature} 
\end{equation}
with $r_{i0} < r_{i1} <  ... < r_{in}$.
It is necessary to prewhiten the feature vector
based upon a PCA analysis of an ideal gas, 
giving a whitened feature factor $\vec{f}^w_i = \vec{W_0} \vec{f}_i$.  
The principal components are then calculated according to
 \begin{equation}
   \vec{p}_i = \vec{W}\vec{f}^w_i. 
 \end{equation}
The machine learning derived order parameter $P_1$
is defined as 
the normalized average of the absolute value
of the first entry of the principal component vector
 \begin{equation}
   P_1 = \langle |p_1| \rangle/\sqrt{\lambda_1}, 
 \end{equation}
where $\lambda_1$ is the eigenvalue 
associated with the first principal component.
PCA performs best when the variance is high, 
so we analyze feature matrices constructed
with data from multiple timesteps that span all possible transitions
in the system.
For example, 
we calculate the features
every 1000 simulation timesteps
during a period of 10$^6$ simulation steps in order to permit
comparison between
early (transient) and late (steady state) phenomena.

In addition to PCA, we apply several other measures to our system.
To separate the clusters from the surrounding gas, we use
a bond orientational order parameter calculated
\cite{Steinhardt83} 
using the algorithm described in Ref.~\cite{Leocmach17}.  
This algorithm computes 
the local tensorial bond orientational order parameter $q_{6m}$, 
where
\[q_{\ell m}(i) = \frac{1}{N_i}\sum_{j=0}^{N_i} Y_{\ell m}(\theta_{ij},\phi_{ij})\] 
for a disk $i$ interacting with $N_i$ neighbors.
The neighbors $N_i$ are determined by a distance cutoff,
$Y_{\ell m}$ are the spherical harmonics,
$\phi_{ij}=0$ for a 2D system,
and $\theta_{ij}$ is the angle between $\vec{r}_{ij}$ and the horizontal.
From $q_{6m}(i)$ we separate the disks into three populations.
The fraction of disks that are inside crystalline clusters is
\[n_x=N_d^{-1}\sum_i^{N_d} \Theta(q_{6m}(i)-x_c)\Theta(N_i-4),\]
where $x_c$ is a minimum threshold value of $q_{6m}(i)$
and all of the clustered disks are required to have a minimum of
four neighbors.
The fraction of disks that are on the edge of a crystal is
\[n_e=N_d^{-1}\sum_i^{N_d} \Theta(q_{6m}(i)-x_c)\Theta(N_i-2)\Theta(4-N_i),\]
where each disk has either two or three neighbors.
Finally, the remaining particles are classified as being in
the gas state,
$n_g=1-n_x-n_e$.
We performed a similar analysis using a Voronoi tessellation to 
identify the gas and edge disks based on the Voronoi cell area, 
and obtained similar results.

\section{Cluster Formation and Evolution}

\begin{figure}
\includegraphics[width=0.45\textwidth]{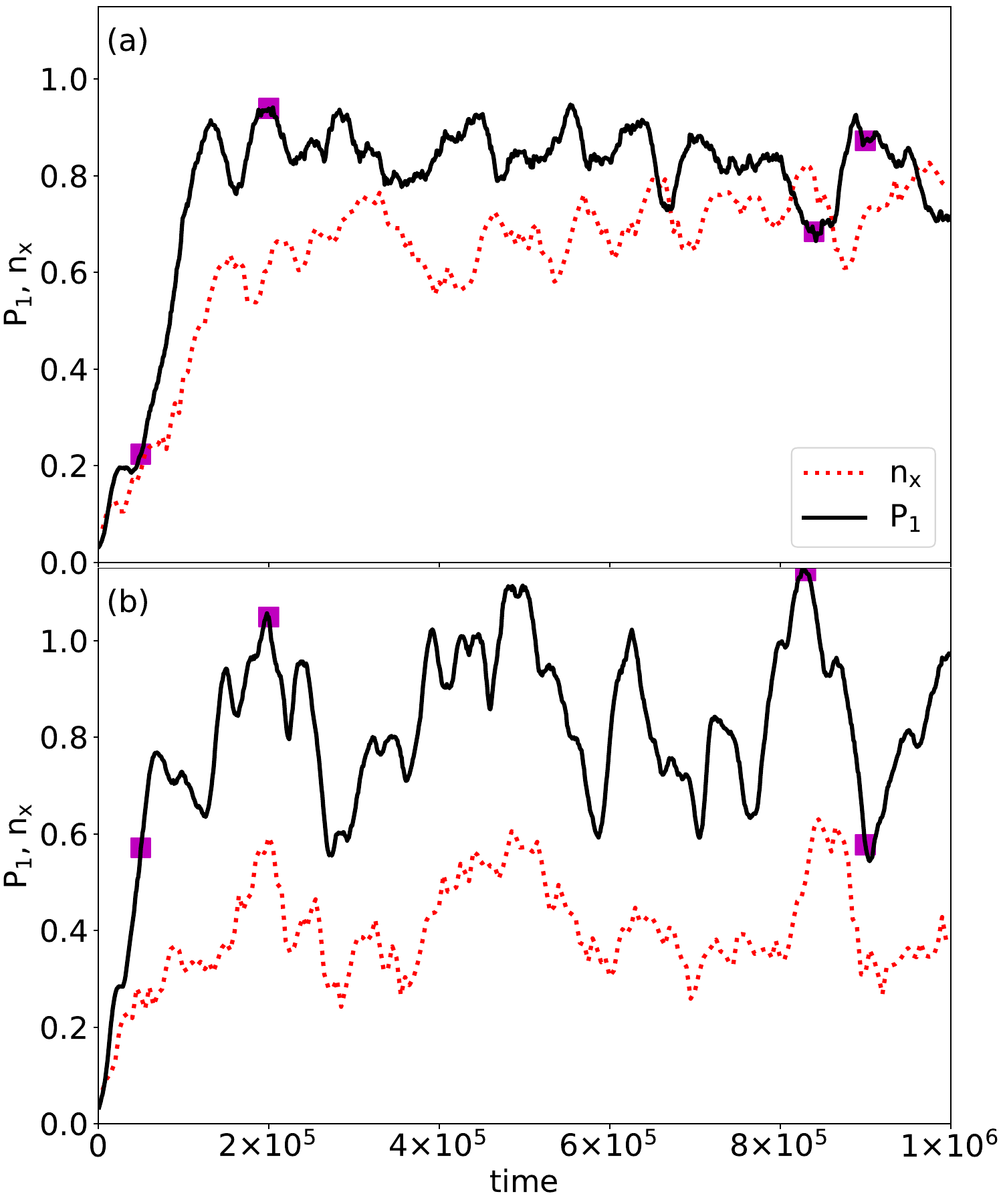}
\caption{
Structural order measurements $n_x$ (red), the fraction of disks located in the
bulk of crystalline clusters, and $P_1$ (black), a machine-learning derived
order parameter, vs time in simulation time steps for samples
with $\phi=0.47$. The persistence time is
(a) $\tau=1 \times 10^5$ and
(b) $\tau=1 \times 10^6$.
Purple squares indicate the times at which the disk images
in Fig.~\ref{fig:1n} were obtained.
} 
  \label{fig:2n}
\end{figure}

\begin{figure*}
\includegraphics[width=0.9\textwidth]{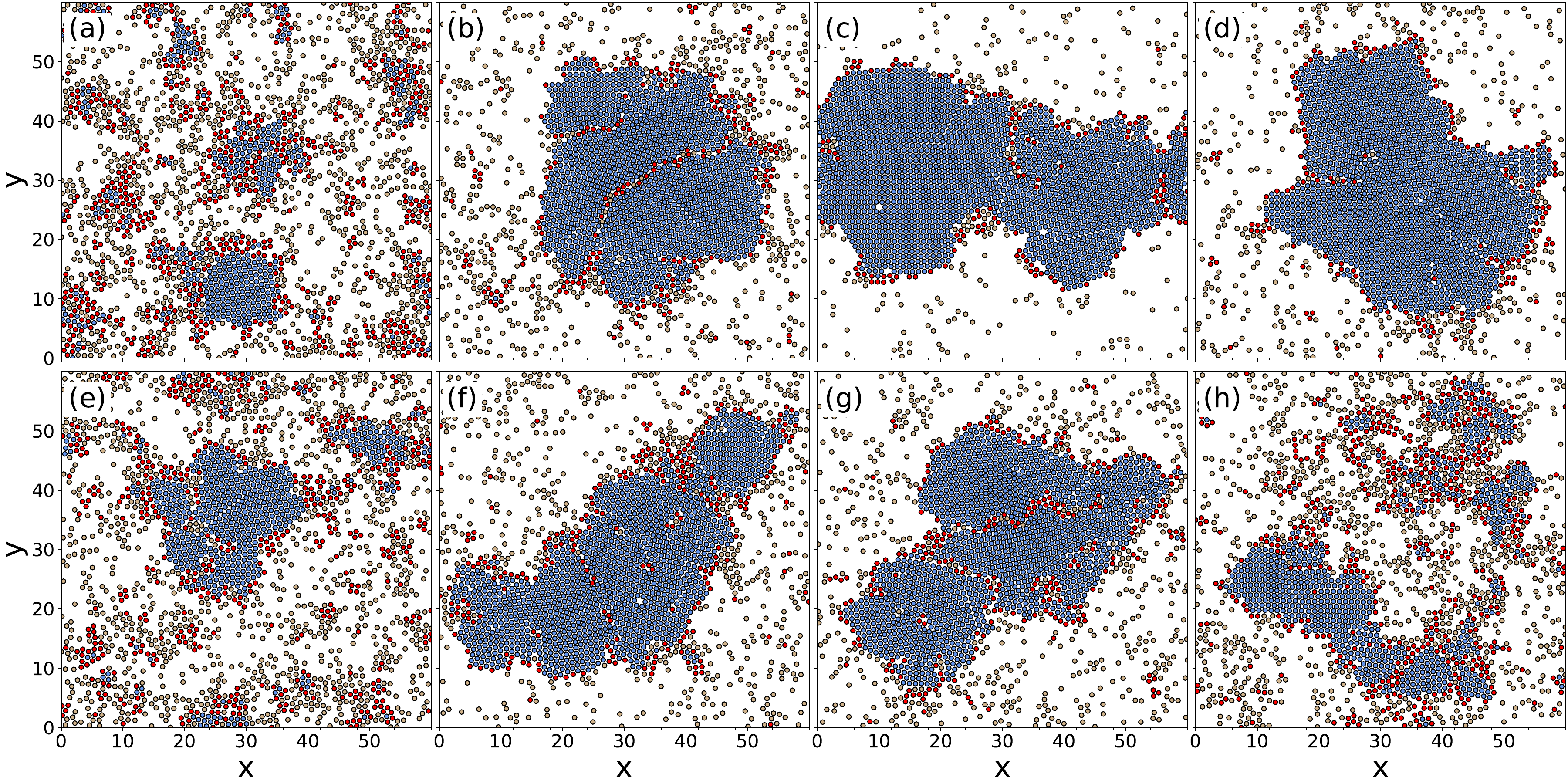}
\caption{
Time evolution of two simulations of actively propelling disks
at a density of $\phi=0.47$ for  
(a-d) $\tau = 1 \times 10^5$ and (e-h) $\tau=1 \times 10^6$.
The systems are shown at times of
(a) $5\times 10^4$,
(b) $2\times 10^5$,
(c) $8.3 \times 10^5$,
(d) $9 \times 10^5$,
(e) $5\times 10^4$,
(f) $2\times 10^5$,
(g) $8.4 \times 10^5$,
and (h) $9 \times 10^5$.
These times are marked with purple squares in Fig~\ref{fig:2n}.
The disks are colored
according to whether they belong to the crystalline fraction
$n_x$ (blue), the gaseous fraction $n_g$ (tan), or
the cluster edges $n_e$ (red)
as determined by the bond orientational order parameter
measurement.
} 
\label{fig:1n}
\end{figure*}

In Fig.~\ref{fig:2n}(a) we plot the structural
order measures $n_x$ and $P_1$
as a function of time
for a system with  $\phi = 0.47$ at a persistence time of
$\tau = 1 \times 10^5$,
and in Fig.~\ref{fig:2n}(b) we show the corresponding
measures for a sample with the same density and $\tau=1 \times 10^6$.
Since the simulations are initialized with the disks
in random, non-overlapping positions, there is no initial structural order,
and both $n_x$ and $P_1$ are low at early times.
The system quickly assembles into a MIPS state,
with a single large cluster forming
within about $t = 1 \times 10^5$ simulation time steps.
For the sample with $\tau = 1 \times 10^5$,
both $P_{1}$ and and $n_x$ grow rapidly at early times
and reach a steady state
with $n_x \sim 0.7$ for $t > 1 \times 10^5$.
These quantities then
fluctuate by approximately 10\%
throughout the remainder of the simulation.
When $\tau = 1 \times 10^6$,
the overall behavior is similar but the
fluctuations in the steady state are more pronounced.
Additionally, the steady state value of $n_x$ for
$\tau = 1 \times 10^6$ is only $n_x \sim 0.41$,
indicating that the clusters
are less stable for the larger value of $\tau$.
In both samples we find that peaks and valleys in $P_1$ and $n_x$ are
correlated with each other, and this behavior is most clearly visible
in the sample with
$\tau = 1 \times 10^6$.

In Fig.~\ref{fig:1n} we illustrate the positions
of the disks in a time sequence
extracted from the time series of
Fig.~\ref{fig:2n} in order to determine whether
the peaks and valleys
occurring in $P_1$ and $n_x$ in Fig.~\ref{fig:2n} are
correlated with structural changes in the system.
Figure~\ref{fig:1n}(a-d) shows images from the sample with
$\tau = 1 \times 10^5$
and Fig.~\ref{fig:1n}(e-h) is for the
sample with $\tau = 1 \times 10^6$.
The disks are colored according to whether they belong to the
crystalline $n_x$ subset, the gaseous $n_g$ subset, or the
edge $n_e$ subset according to the bond orientational order
parameter measurement.
At early times,
as shown for $t = 5\times10^4$
in Fig.~\ref{fig:1n}(a,e), 
only small clusters
have evolved from the initial disordered state of the system, indicating that
smaller clusters
are associated with small values of $P_{1}$ and $n_{x}$.
The ML algorithm is able to discern the average cluster size,
with $P_1$ = 0.25 for $\tau = 1 \times 10^5$ where the incipient crystalites
are small, but $P_1$ = 0.55 for
$\tau = 1 \times 10^6$ where the incipient crystalites are larger.
While both $P_1$ and $n_x$ make use of
interparticle distances and local structural information,
$P_1$ in general performs better than $n_x$ in detecting crystalline signatures
since it is able to capture ordering that
extends beyond the nearest neighbors.

Figure~\ref{fig:1n}(b,f) shows that for both
values of $\tau$, a large active cluster
forms at $t = 2\times 10^5$.
Here $P_{1} \approx 1.0$.
At longer times in the $\tau = 1 \times 10^5$ sample,
both $P_1$ and n$_{x}$
remain large
with $P_1 \approx 0.9$ and $n_x \approx 0.7$, indicating
that the cluster size reaches a steady state.
This is consistent with the stable large cluster visible in
Figs.~\ref{fig:1n}(c) and (d).
The morphology of the cluster continues to evolve
as a function of time, with grain boundaries forming and then
disappearing,
but the overall size of the cluster remains nearly constant
once it has formed.
In contrast, although a large cluster is still present in
the $\tau = 1 \times 10^6$ system
at $t = 8.4\times 10^5$, as shown
in Fig.~\ref{fig:1n}(g),
this is followed by the rapid disassembly event illustrated
in Fig.~\ref{fig:1n}(h).
When the cluster disassembles,
$P_{1}$ drops to $P_1 \approx 0.6$ and
$n_x$ also shows a less pronounced drop.
In the supplementary
information
we include a video illustrating
the disassembly of the cluster for this time interval \cite{Suppl}.
As indicated by the recovery of the value of $P_1$ at later times in
Fig.~\ref{fig:2n}(b), the cluster repeatedly reforms and disassembles
for this value of $\tau$. Thus, although there is clustering for
both $\tau=1 \times 10^5$ and $\tau = 1 \times 10^6$, the
$\tau=1 \times 10^5$ cluster is stable while the
$\tau=1 \times 10^6$ cluster is strongly fluctuating.

\begin{figure}
\includegraphics[width=0.45\textwidth]{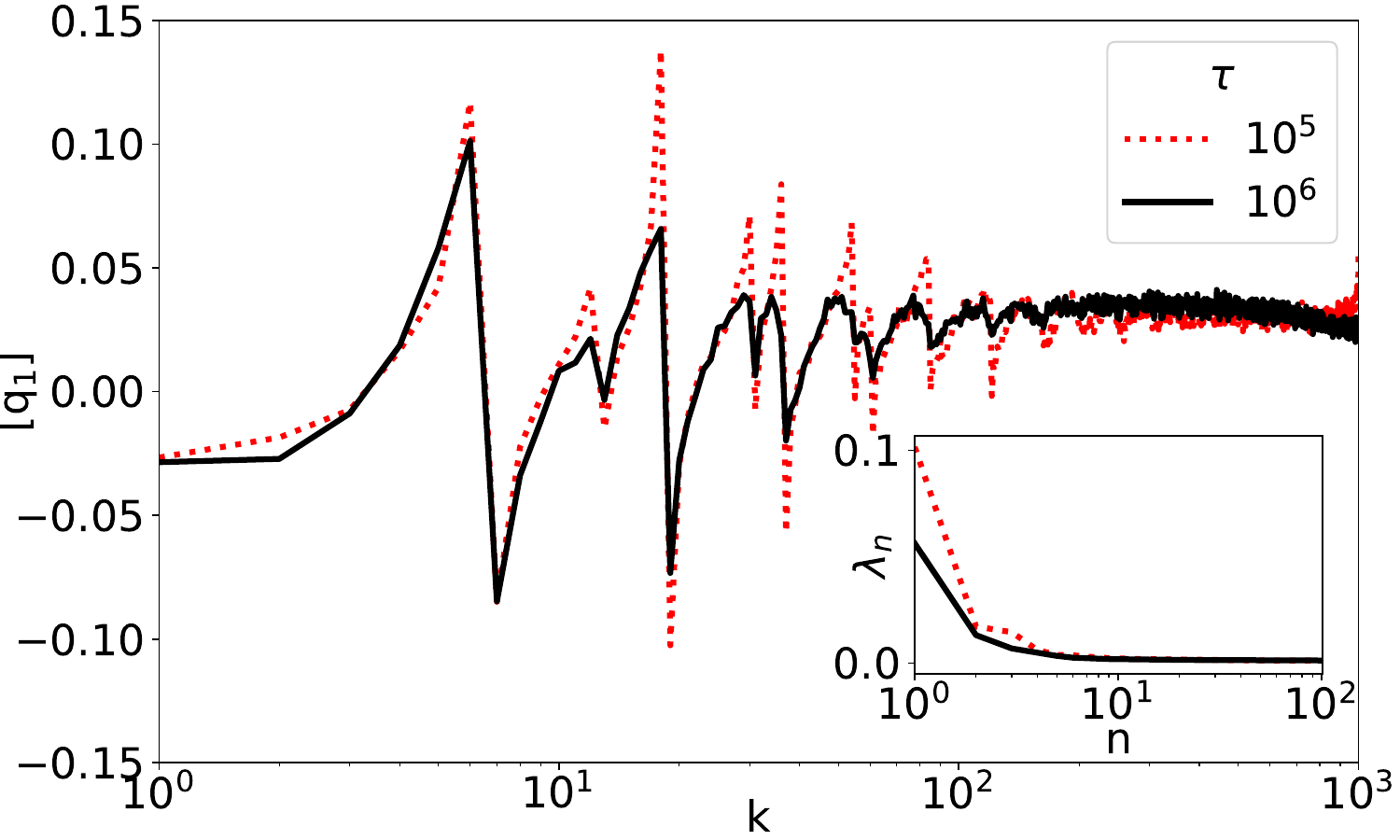}
\caption{
$[q_1]$, the first row of the total transformation matrix
$\vec Q=\vec W\vec W_0$, vs component number $k$ for the system in
Figs.~\ref{fig:2n} and \ref{fig:1n} with $\phi=0.47$ at
$\tau=1 \times 10^5$ (red) and $\tau = 1 \times 10^6$ (black).
This measure illustrates how PCA effectively weights $g(r)$.
Inset:
The
normalized PCA eigenvalues $\lambda_n$ vs $n$ for the same samples.
} 
\label{fig:3n}
\end{figure}

Both
$P_1$ and $q_{6m}$
are affected by 
the number of particles participating in the largest
active cluster and the local ordering of particles.
In general, $P_1$ captures more information than $q_{6m}$.
To demonstrate this,
in Fig.~\ref{fig:3n} we plot the metrics derived from PCA
for the
entire time series obtained for the
systems in Fig.~\ref{fig:2n} and Fig.~\ref{fig:1n}
with $\tau = 1 \times 10^5$
and $\tau = 1 \times 10^6$.
The main panel
of Fig.~\ref{fig:3n}
is a plot of the components of $[q_1]$,
the first row of the total transformation matrix
$\vec Q=\vec W\vec W_0$, versus component number $k$.
Since the $\vec Q$ metric contains a geometric snapshot of
the neighbor environment for many probe particles,
the quantity $[q_1]$ provides information similar to the radial
distribution function $g(r)$, with increasing $k$ corresponding
to increasing values of $r$ due to the fact that the feature
vectors $\vec f$ have been sorted according to size.
The PCA effectively performs a weighting of $g(r)$, as
described in more detail in Ref.~\cite{McDermott20}, and
in Fig.~\ref{fig:3n} we observe that the weightings differ when
$\tau$ is varied. At $\tau=1 \times 10^5$, where stable large
clusters form, $[q_1]$ is strongly peaked at the regular spacings
expected for a crystalline lattice out to relatively large
distances, indicating that PCA has detected longer range crystalline
ordering. In contrast, for $\tau=1 \times 10^6$, the weight at these
crystalline peaks is diminished and is replaced by greater weight
at a continuum of longer distances, representative of the increased
amount of gas phase present.
The inset of Fig.~\ref{fig:3n} shows the normalized eigenvalues
$\lambda_n$ sorted by magnitude, corresponding
to the ranked
principal components plotted as a function of element index $n$.
This is a typical performance measurement of PCA's ability to
reduce the dimensionality of the feature space.
Here the rapid decrease in $\lambda_n$ with $n$ demonstrates
that the first few principal components are able to capture
most of the information in the data set.
The ratio of $\lambda_1$ to $\lambda_2$ is higher
for
$\tau = 1 \times 10^5$ than for $1 \times 10^6$,
indicating
that more of the feature space can be described by the first
principal component in the $\tau = 1 \times 10^5$ system. This is
consistent with the enhanced amount of crystalline ordering that appears
at $\tau = 1 \times 10^5$.

\begin{figure}
 \includegraphics[width=0.45\textwidth]{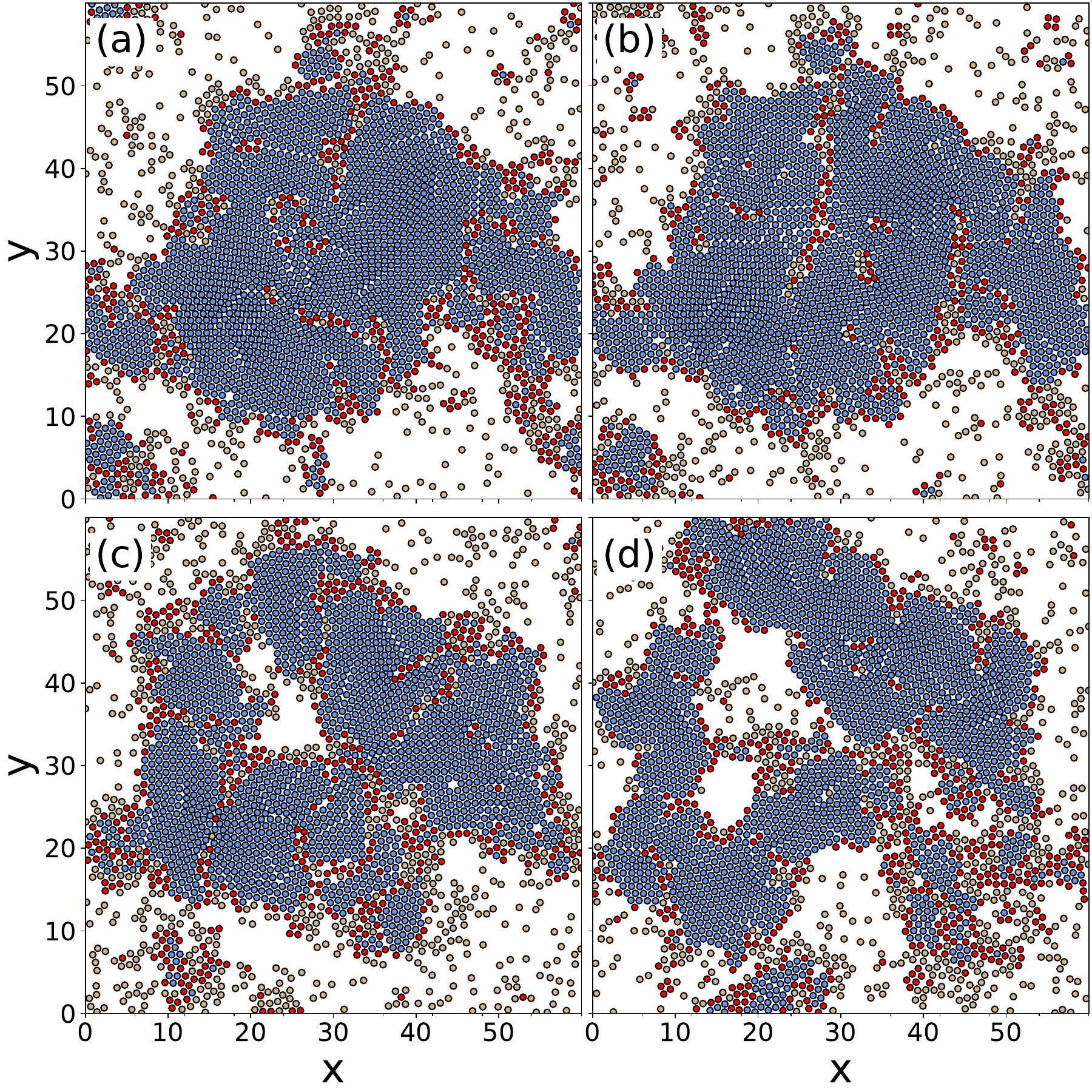}
 \includegraphics[width=0.45\textwidth]{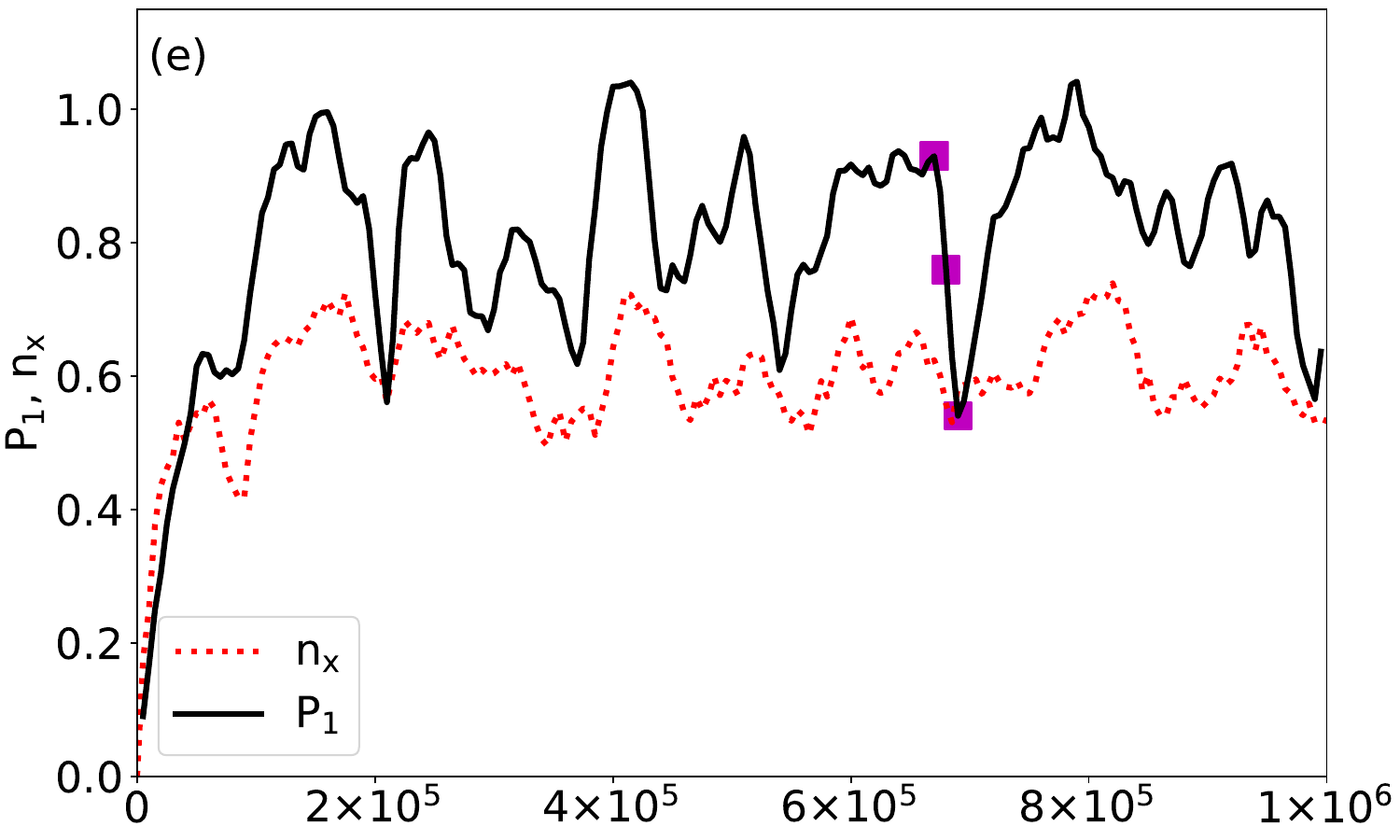}
\caption{
(a-d) Illustration of the time evolution of the formation of a void
in a sample with $\phi=0.6$ and $\tau=1 \times 10^5$.  
The system is shown at times of
(a) $6.7 \times 10^5$,
(b) $6.72 \times 10^5$,
(c) $6.8 \times 10^5$, and
(d) $7.0 \times 10^5$.
The disks are colored according to whether they belong to $n_x$ (blue),
$n_g$ (tan), or $n_e$ (red). An animation of this time sequence is available
in the supplemental material \cite{Suppl}.
(e) Time series of $P_1$ (black) and n$_{x}$ (red) for the
same system. 
Times
$6.7 \times 10^5$,
$6.8 \times 10^5$,
and $7.0 \times 10^5$,
corresponding to the interval illustrated in panels (a-d),
are highlighted with purple squares.
We note that the times of panels (a) and (b) are so close together that
they overlap on the scale of the squares, so only the time of panel (a) is
shown.
} 
\label{fig:8n}
\end{figure}

To show more clearly how
time-resolved $P_1$ values correlate with the actual dynamics
of the system, in
Fig.~\ref{fig:8n}(a-d) we
illustrate the particle positions during the destruction of a large
cluster due to a void formation process
for a system with $\phi=0.6$ and $\tau = 1 \times 10^6$.
The fragments of grain boundaries visible as chains of red particles
in Fig.~\ref{fig:8n}(a) coalesce into a prominent longer and thicker grain
boundary in Fig.~\ref{fig:8n}(b).
The evolution of the grain boundaries is controlled in part by the
one-dimensional
string-like motion of vacancies through the crystalline regions
\cite{Pertsinidis01,Dudarev03,Smallenburg12,Lechner13}, visible
as white streaks in the figure. A vacancy cannot remain stable inside the
crystalline region but travels quite rapidly toward vacancy sink areas
\cite{Swanson82}
consisting of grain boundaries and the edges of the cluster.
The vacancies can move at speeds considerably higher than the motion of
a free individual disk, and their movement also becomes increasingly
one-dimensional as the
density of the system becomes higher \cite{vanderMeer18}.
In Fig.~\ref{fig:8n}(c), continued absorption of vacancies by the grain
boundary results in the opening of a void area that serves
as a vacancy sink location,
and the effectiveness of this sink
increases as the
outer contour of the void becomes longer.
Further growth of the void causes the cluster
to break apart, as illustrated in
Fig.~\ref{fig:8n}(d).
We find that when $\tau = 1 \times 10^5$, vacancies are short lived and
tend to be reabsorbed by the crystal almost as quickly as they form,
while when 
$\tau = 1 \times 10^6$, the vacancies are able to persist for much
longer times and can travel distances on the order of the size of the
cluster.
This can also be seen in the supplemental movies \cite{Suppl}.

In Fig.~\ref{fig:8n}(e), we plot $n_x$ and $P_1$ versus time for the
$\phi=0.6$ and $\tau=1 \times 10^6$ system illustrated
in Fig.~\ref{fig:8n}(a-d).
The purple squares correspond to the times at which images of the
void formation and growth were obtained. 
Although there is a weak dip in $n_x$ over this time interval, it is
difficult to tell from $n_x$ that anything unusual is happening to
the cluster structure. In contrast, $P_1$ exhibits a pronounced
dip as the cluster disintegrates.
When the cluster is still largely intact in
Fig.~\ref{fig:8n}(a,b), $P_1$ has a high value of $P_1\approx 0.9$,
but when the cluster splits apart, $P_1$ drops to
$P_1 \approx 0.59$.
The reassembly of the cluster (not illustrated) at slightly later times
is accompanied by a recovery of the value of $P_1$ back to
its former level of $P_1 \approx 0.9$.
This shows
that $P_{1}$ captures many more features than
are found in standard measures
used to characterize MIPS, and that even the time evolution of $P_1$
contains useful information.

\section{Velocity Fluctuations and Noise Power}

Since we have obtained a time-dependent measure capable of capturing
the evolution of the MIPS structure in the steady state, it is interesting
to examine the time series of related fluctuating quantities in more
detail.
For many systems such as magnetic domain walls \cite{Sethna01,Durin05},
charge density waves, \cite{Gruner88},
Wigner crystals \cite{Reichhardt04}, and
vortices in type-II superconductors
\cite{Marley95,Field95,Olson98a,Okuma08,Reichhardt17},
under application of an external drive it is possible to analyze
fluctuations in the velocity or resistance for different conditions.
In the active matter system,
the net velocity is zero since there is no net bias in the run-and-tumble
motion and on average the particles move in all directions at once;
however, we can
still measure the average speed of all the particles as
a function of time since this quantity remains finite.
We compute the average speed $\langle V\rangle$ according to:
\begin{equation}
  \langle V \rangle = \frac{1}{N_d}\sum_i^{N_p} \sqrt{(\vec{v}_i\cdot {\hat x})^2 +  
    (\vec{v}_i \cdot {\hat y})^2}   
\end{equation}
where the average is taken over all particles at an instant in time.

\begin{figure}
\includegraphics[width=0.49\textwidth]{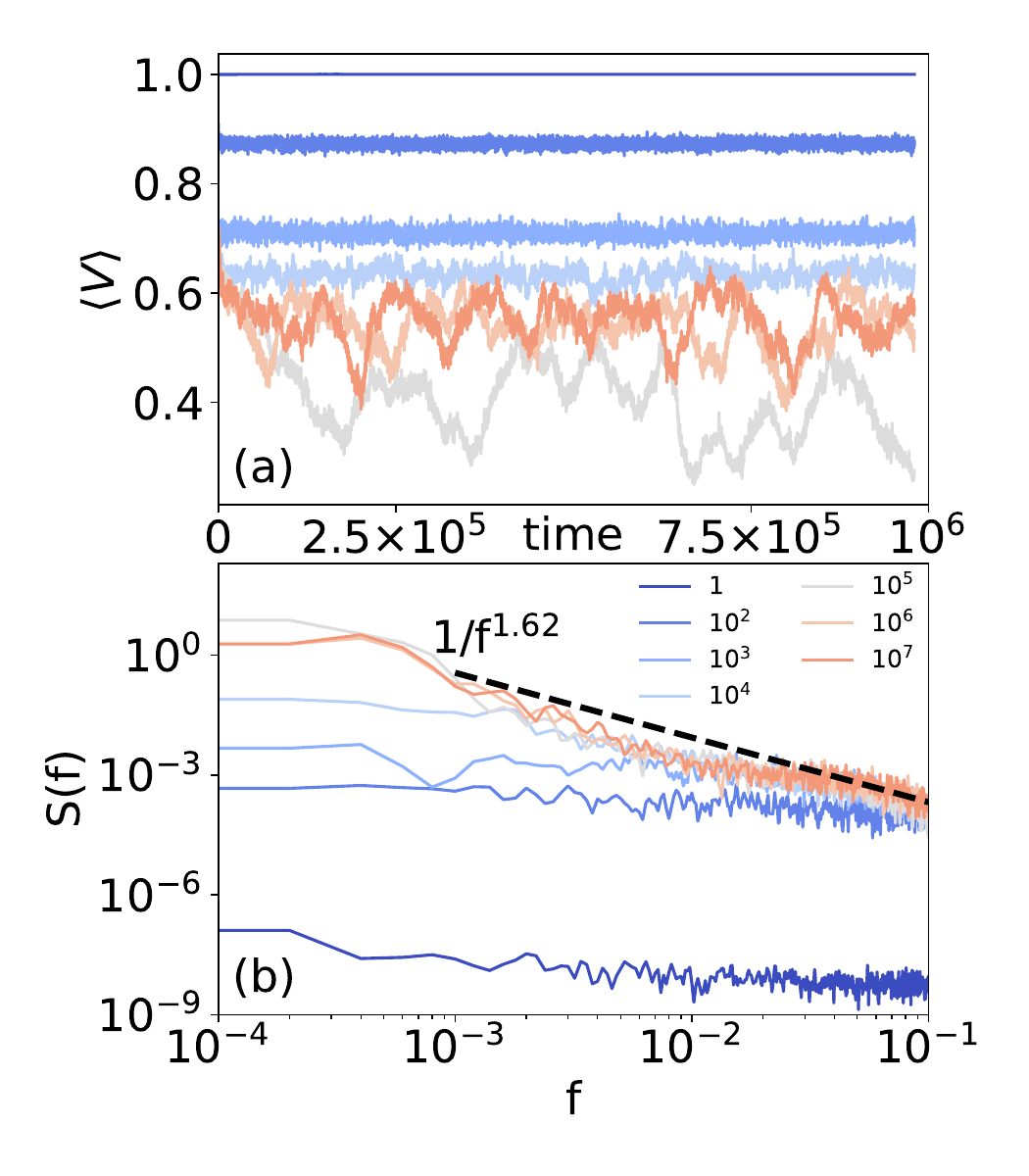} 
\caption{
(a) Average speed $\langle V \rangle$ vs time for samples
with $\phi=0.45$ and $\tau=1$, 100, $10^3$, $10^4$, $10^5$, $10^6$, and $10^7$,
from top to bottom. The curves are colored according to the legend in panel (b).
(b) Noise spectra $S(f)$ obtained from the time series in panel (a).
The dashed line indicates a fit to $S(f) \propto 1/f^{\alpha}$ with
$\alpha=1.62$.
} 
\label{fig:4n}
\end{figure}

In Fig.~\ref{fig:4n}(a)
we plot $\langle V\rangle$ vs time for systems with $\phi = 0.45$
at $\tau = 1 \times 10^0$,
$1 \times 10^2$, $1 \times 10^3$, $1 \times 10^4$,
$1 \times 10^5$, $1 \times 10^6$, and $1 \times 10^7$.
For $1 < \tau < 1 \times 10^4$, the system is in a fluid phase
where $\langle V\rangle$ fluctuates weakly and the magnitude
of $\langle V\rangle$ decreases monotonically with increasing $\tau$.
When $\tau = 1$,
every particle tumbles at every timestep so the persistence of the
motion drops to zero and the particle motion is unimpeded,
giving $\langle V\rangle = F^a = 1$.
As $\tau$ increases in the fluid state,
more collisions occur between particles,
causing them to slow and decreasing the
average speed.
Some small clusters begin to form
for $\tau = 1 \times 10^4$, leading to an increase in the characteristic
time scale of fluctuations
in $\langle V\rangle$.
At $\tau = 1 \times 10^5$ the system has
entered the MIPS phase,
where there are large fluctuations in
$\langle V\rangle$ and we observe formation of a large cluster.
For $\tau = 1 \times 10^6$ and $1 \times 10^7$,
the system is still in
a MIPS state but
the fluctuations of $\langle V\rangle$ are diminished
and the magnitude of $\langle V\rangle$ has increased compared
to the $\tau = 1\times 10^5$ system, consistent with
the clusters being more stable at
$\tau=1 \times 10^5$ than at $\tau=1 \times 10^6$,
as described in Section IV.

The time series of $\langle V\rangle$
can be characterized with the power spectrum
\begin{equation}
\label{eq:fft}
S(f) = \left|\int \langle V(t) \rangle \exp{(i 2\pi f t)} dt\right|^2  \ ,
\end{equation}
where $f=2\pi\omega$ is the frequency.
Note that this measure differs from the density fluctuations
examined in previous work \cite{Fily12}.
In Fig.~\ref{fig:4n}(b) we plot $S(f)$
for the time series from 
Fig.~\ref{fig:4n}(a).
For $\tau = 1$,
$S(f)$ is nearly flat
across six decades of frequency,
indicative of white noise.
As $\tau$ increases, we find a transition from 
white noise to a power law spectrum with
  \begin{equation}
  \label{eq:powerlaw}
          S(f) \propto f^{-\alpha} \ ,
  \end{equation}
where the amount of noise power at low frequencies increases significantly
in the power law regime.
For $\tau = 1 \times 10^5$
we obtain a fit to $\alpha = 1.62$ using
the algorithm in Ref.~\cite{Alstott14}.
We find similar values of $\alpha$
for $\tau = 1 \times 10^6$ and $1 \times 10^7$, although
both the range of the power law
and the magnitude of the low frequency noise power are greatest at
$\tau = 1 \times 10^5$.

Transitions from white noise to $1/f^{\alpha}$ noise
accompanied by peaks in
the low frequency noise power are observed
in many driven systems that exhibit depinning \cite{Reichhardt17}.
In experiments on the
depinning of superconducting vortices,
Marley {\it et al}. \cite{Marley95} found exponents ranging from
$\alpha = 1.5$ to $\alpha = 2.0$ in the peak effect region,
where the motion of the vortices is expected to be plastic.
In studies of vortex avalanches, exponents of $\alpha=1.5$ were
observed in the power law distribution of avalanche sizes for
drives just above depinning, while $\alpha$ increased to $\alpha=2.0$ as
the driving rate was increased \cite{Field95}.
In simulations of
the depinning of  2D vortex assemblies,
$\alpha \approx 1.5$  to $1.9$ depending on the strength of the
quenched disorder in the sample
\cite{Olson97,Olson98a,Reichhardt17}.
Simulations of yielding transitions in 2D amorphous solids undergoing
avalanches have produced
power law dependences with $\alpha = 1.25-1.35$ \cite{Regev19}.

Power law behavior arises near a critical point because the diverging
length scales that emerge lead to larger and larger amounts of
low frequency noise.
For the nonequilibrium random field Ising model near a
disorder-induced nonequilibrium critical point,
noise power spectra with $\alpha = 1.77$
are observed \cite{Travesset02},
and the relationship between the value of $\alpha$ and the critical
exponents associated with the transition is given by 
$\alpha = 1/\sigma\nu z$.
Chen and Yu \cite{Chen07}
examined noise fluctuations in Ising and Potts models
and found that the noise power
of energy fluctuations reaches a maximum
at the magnetic phase transition,
with white noise
appearing above and below the transition.
For the Ising model,
$\alpha = 1.0$, while for the Potts model,
$\alpha = 1.56$.
If magnetization fluctuations are considered instead of energy
fluctuations, power law noise still arises with
$\alpha = 1.8$ for the Ising model and $\alpha = 1.7$ for the Potts model.

\section{$\phi-\tau$ Heat Diagrams}

\begin{figure}[t]
\includegraphics[width=0.49\textwidth]{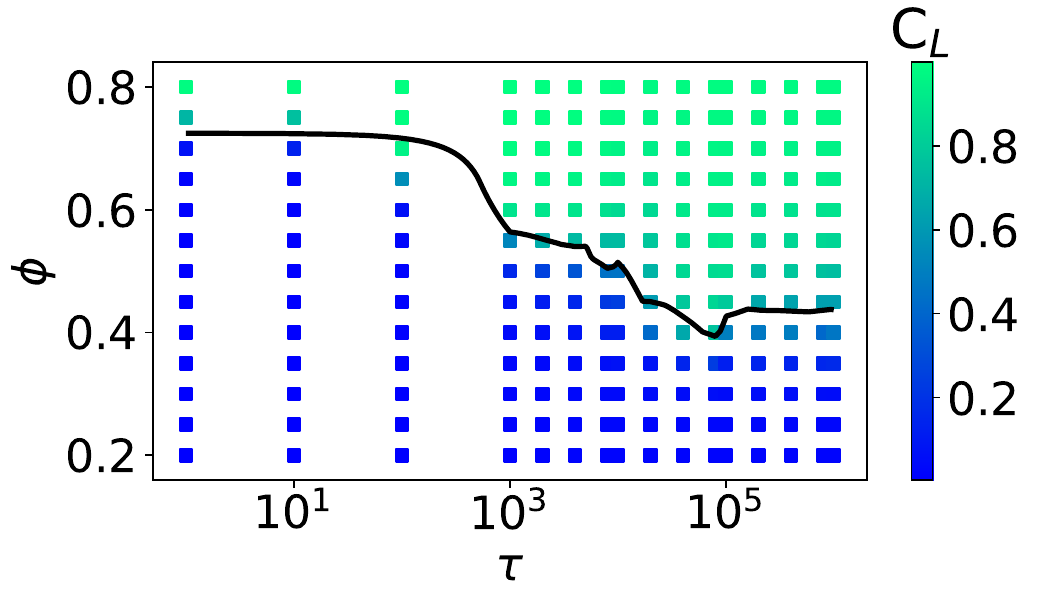}
\caption{
(b) Heat map of $C_L$, the fraction of particles in the largest
cluster, as a function of $\phi$ vs $\tau$. The black line indicates
points where $C_L=0.5$.  
}
\label{fig:CL}
\end{figure}

One of the most common measures of motility-induced phase separation is
the fraction of particles $C_L$ contained in the largest connected cluster.
This measure is similar to $n_x$ but
does not distinguish whether the arrangement of the particles is
crystalline; it only determines whether the particles are in contact
with each other.
In Fig.~\ref{fig:CL}, we plot a heat map of
$C_{L}$ as a function of $\phi$ vs $\tau$; the black line
indicates points where $C_L=0.5$.
We observe MIPS phases for $\tau \geq 1 \times 10^3$.
For $\tau < 1 \times 10^3$ and $\phi> 0.7$, the system is sufficiently
dense that the value of $C_L$ is large, but this is the result of a
percolating network of particle-particle contacts rather than the
formation of a MIPS cluster.
For
$0.4 < \phi < 0.5$ and
$1 \times 10^4 < \tau < 1 \times 10^6$,
$C_L$ exhibits reentrance as a function of $\tau$.
Since $C_{L}$ can
only distinguish between
the presence or absence of extended networks of particle contacts,
and cannot distinguish a cluster from a percolating contact network,
it is necessary to turn to other measures to determine whether
additional subphases exist within the MIPS state, such as
the stable and unstable clusters described in Section IV for different
values of $\tau$.

\begin{figure}[t]
\includegraphics[width=0.49\textwidth]{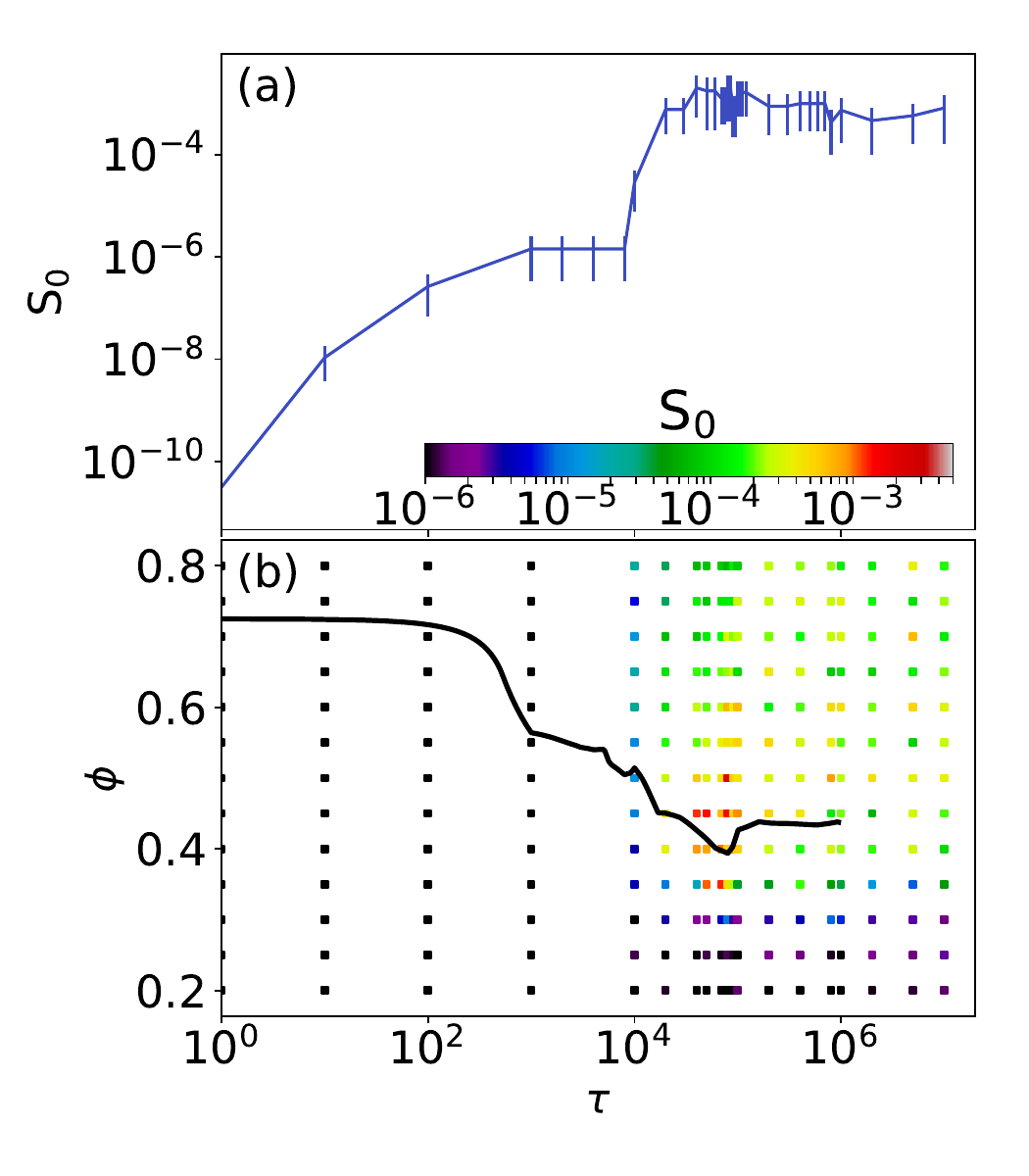}
\caption{
(a) Low frequency noise power $S_0$ versus $\tau$ for samples with
$\phi=0.45$.
(b) Heat map of $S_0$ as a function of $\phi$ vs $\tau$. Black points
indicate denote $(\phi,\tau)$ pairs for which
the noise is white and $S_0$ has a very low value.
}
  \label{fig:5n}
 \end{figure}

In Fig.~\ref{fig:5n}(a) we 
plot the noise power $S_{0}$
versus $\tau$ for a sample with $\phi = 0.45$.
To obtain the noise power,
we integrate $S(f)$ over a small interval at low frequencies,
\begin{equation}
\label{eq:S0}
  S_0 = \int_{f_0}^{f_1} S(f) df \ .
\end{equation}
We average the measurement of $S_0$
over several intervals $(f_0,f_1)$ 
and report the measure with an associated error. 
We find that $S_0$ increases steadily from $\tau = 1$, 
with a sharp increase
occurring near $\tau = 1 \times 10^4$.
The noise power reaches
a peak 
near $\tau = 5 \times 10^4$,
and stabilizes onto a plateau for
$\tau > 1 \times 10^5$.

Figure~\ref{fig:5n}(b) shows a heat map of $S_0$
plotted as a function of $\phi$ versus $\tau$.
In the fluid phase,
the noise power is low and the power spectra are white.
For $\phi > 0.75$ and $\tau < 1 \times 10^3$, the noise is low
and the power spectra are white even though $C_{L}$ is high,
indicating that the system is not in the MIPS
state.
We find that $S_0$ is nonmonotonic and reaches its highest values
for $5\times 10^4 \leq \tau \leq  5\times 10^5$
and $0.35 < \phi < 0.55$,
which is also the region where $C_L$ becomes reentrant
as a function of $\tau$.
In this regime, the velocity noise has
a $1/f^\alpha$ signature
with $\alpha \approx 1.6$.
Elsewhere in the MIPS regime, the velocity noise still has
a $1/f^\alpha$ tail but the value of $S(f)$ saturates
at low frequencies,
whereas in the critical regime where $S_0$ is largest,
the power law behavior of the velocity noise extends over the
greatest range of frequencies.
The black line indicates the points at which
$C_{L} = 0.5$, showing that more information can be extracted from  
the behavior of $S_0$ than from the behavior of $C_L$ alone.
In particular, the noise measurement captures some features outside
of the boundary delineated by the $C_L$ measurement. Near
$\tau = 1\times 10^5$ there is large noise all the way down to
$\phi=0.3$, while the cluster measurement was only large down to
$\phi=0.4$. At the low density of $\phi=0.3$, there are short-lived
loosely packed clusters in which the particle speed drops, but these
clusters are not compact enough to register in the $C_L$ measurement.

The results in Fig.~\ref{fig:5n}(a)
are similar to the noise power measurements obtained in
driven systems, where a maximum in the noise power occurs at
a dynamical transition and is then followed by a decrease in noise.
Chen and Yu found that
the noise power peaks at the critical point
for equilibrium phase transitions \cite{Chen07}.
For driven systems with quenched disorder that undergo plastic
depinning, such as
superconducting vortices,
there is a peak in the noise power within the plastic flow
phase where strong distortions of the vortex lattice are occurring.
The noise power decreases when
the system is more strongly driven since the number of plastic
events is reduced \cite{Marley95,Olson98a,Reichhardt17}.
In the active system,
as $\tau$ increases,
the cluster structure becomes more crystalline and fewer plastic
events occur.
For $\phi > 0.55$, the peak in $S_0$ is reduced in magnitude and becomes
more of a plateau.
This suggests that the transition into the MIPS state has
properties associated with a critical phase transition, and that
in the region spanning $0.35 < \phi <  0.6$ these critical properties
are visible, while for larger $\phi$ and higher $\tau$,
the system is not critical.

\begin{figure}
\includegraphics[width=0.45\textwidth]{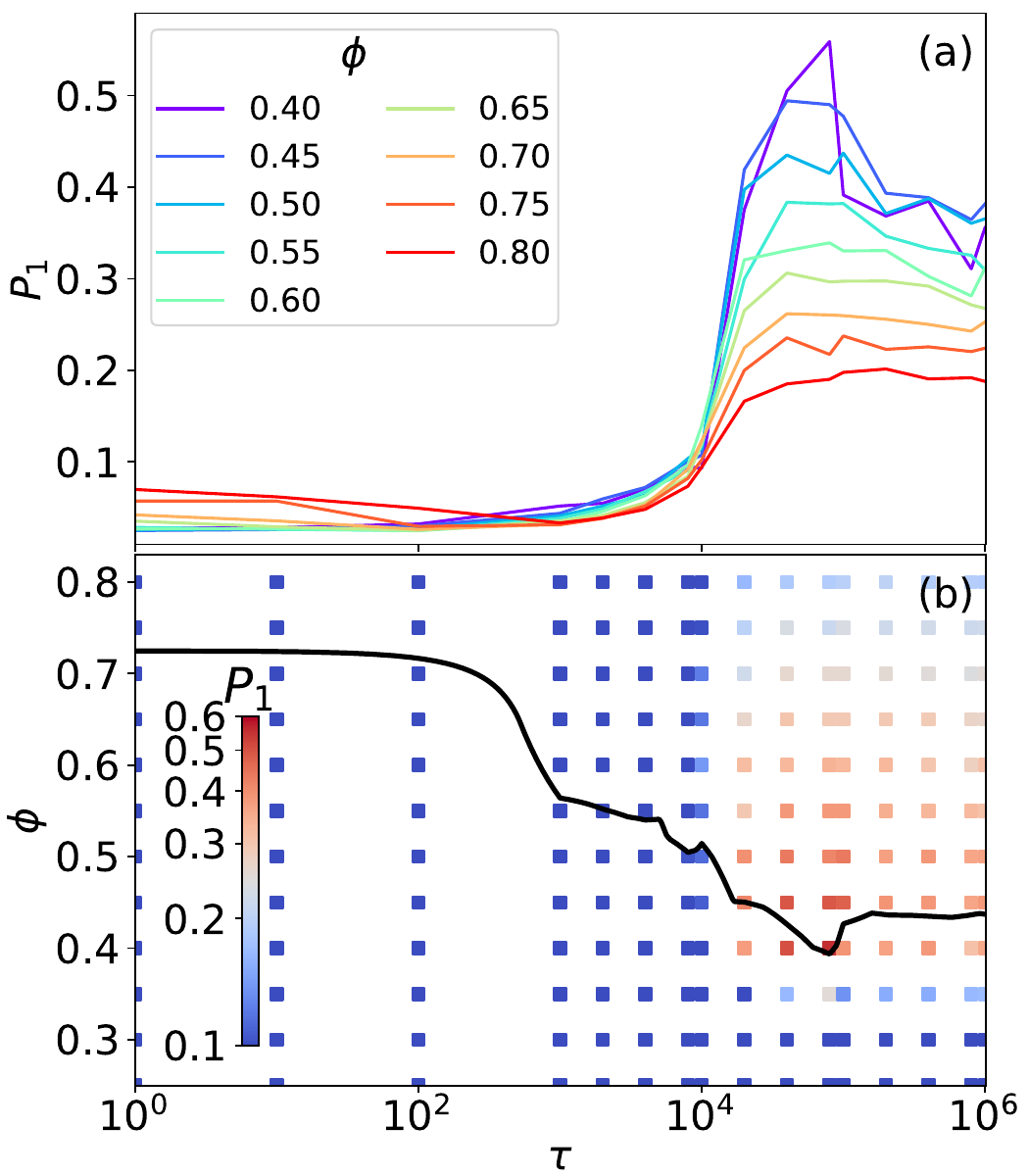} 
\caption{
(a) P$_1$  vs $\tau$ for varied disk density
$\phi=0.4$ (purple), 0.45, 0.5, 0.55, 0.6, 0.65, 0.7, 0.75, and 0.8 (red),
from top
to bottom.
(b) Heat map of $P_1$ as a function of $\phi$  vs $\tau$.
The black line indicates points at which
$C_L =$  0.5.
This figure was composed using
500 GB of feature data extracted from 90 simulations.
} 
\label{fig:8}
\end{figure}

We next measure $P_1$ for varied $\tau$ and $\phi$ to
see whether the ML detects features that are similar or different
from what appears in $C_L$ and $S_0$.
In Fig.~\ref{fig:8}(a) we plot P$_1$ versus $\tau$ for
samples with $\phi=0.4$, 0.45, 0.5, 0.55, 0.6, 0.65, 0.7, 0.75, and 0.8.
Here,
feature data are evaluated every $1 \times 10^5$ timesteps
across 10$^6$ simulation steps and combined into a single matrix
for evaluation of principal components. For
$\tau > 10^6$,
it becomes difficult to analyze the data.
In the liquid phase, $P_1$ is low
even for $\phi = 0.8$ where $C_{L}$ is high.
For $0.4 < \phi  < 0.55$ in the lower density
portion of the MIPS regime, there is a peak in $P_{1}$
that also coincides
with the peak in $S_{0}$.
For
$\phi > 0.55$, $P_{1}$ decreases
and saturates to a level higher than its value
in the liquid phase, since MIPS is still present.
In Fig.~\ref{fig:8}(b), we plot a heat map of $P_{1}$
as a function of $\phi$ versus $\tau$ along with
the boundary line indicating the points where
$C_{L}=0.5$.
The structure of $P_1$ is similar to that of $S_0$, with the largest
values of $P_1$ occurring in the low density, intermediate $\tau$ portion
of the MIPS regime.

\section{Discussion}

\begin{figure}
\includegraphics[width=0.45\textwidth]{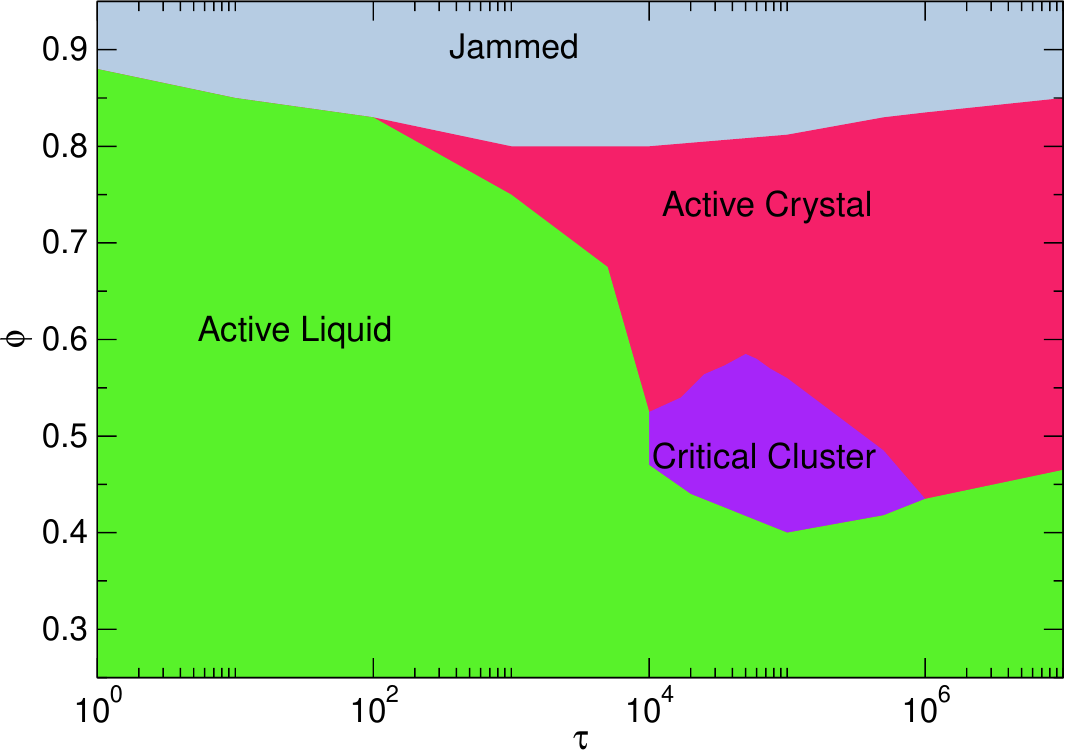}
\caption{
A schematic of the possible regimes inside the MIPS state as a
function of $\phi$ vs $\tau$ showing a jammed crystal,
an active crystal, a critical region, and an active liquid. 
}
\label{fig:9}
\end{figure}

The different features in $C_L$, $S_0$, and $P_1$ suggest that there
could be different regimes within the MIPS phase.
For example, at high $\phi$
and high $\tau$, $P_{1}$ and $S_0$ have intermediate values
but $C_{L}$ is high. This regime could be described as
consisting of
active crystals where there are still
fluctuations or intermittent motion
giving rise to avalanche-like rearrangements
inside the crystals, but the size
of these rearrangements is cut off
so that the magnitude of the low-frequency noise is reduced.    
Mandal {\it et al.} \cite{Mandal20}
considered the limit of dense active systems and varied
both the motor force and the persistence time.
At small motor forces,
the system is in a dynamically arrested state but
is still active so that it can be
described as an active crystal.
At high motor forces the system shows intermittent plastic
yielding, while for even higher motor forces, the system becomes
a liquid due to interpenetration of the particles. A similar
breakdown of clustering due to particle interpenetration
was observed earlier \cite{Fily14}.
In our case, we hold the motor force fixed so that particle interpenetration
remains unimportant even as the density increases.
For high density and 
larger values of $\tau$, our system remains
in the active crystal state but becomes
more frozen and exhibits reduced intermittency.
We argue that 
for the dense system, the noise will be small in the active crystal but high
when the motor force becomes large enough to induce plastic deformations.
Such deformations would give rise to a $1/f^\alpha$ velocity
fluctuation signature and
large noise, and the noise
power would decrease again in the active liquid state.
For dense systems with very low persistence length,
the system behaves more like a jammed solid with low noise and low $P_1$.
The critical regime we find where $P_1$ and $S_0$ are large
could be associated with a second-order transition
into the MIPS phase or some other type of critical regime
similar to the criticality associated with
plastic depinning.
The critical regime is characterized by high $C_L$, high $P_{1}$,
and high $S_0$. 
In Fig.~\ref{fig:9}, we plot a suggested schematic phase diagram
as a function of $\phi$ versus $\tau$ showing 
an active liquid, a jammed phase, an active crystal, and a critical
phase.

Our results suggest that the dynamics of vacancies play an
important role in determining whether the large MIPS cluster can
remain stable once it forms in the critical cluster regime,
or whether the cluster will repeatedly
break apart and reform in the active crystal regime.
If we focus on the nucleation of a single vacancy, a nucleation attempt can
succeed if a neighboring particle does not move into the newly formed
vacancy location
\cite{Zhurkov65}. The neighboring particle is confined by the nearly
crystalline surrounding cage of dense phase, and explores different
directions of this cage at a rate of $1/\tau$. When $\tau$ is low
enough, the
neighboring particle changes running directions frequently and can explore
the entire cage rapidly.
The rate of
vacancy discovery decreases with increasing $\tau$, so that for high $\tau$
the ability to fill the vacancy decreases with increasing $\tau$ since the
neighboring particle is unable to find the vacancy efficiently. As a result,
vacancies are filled at
a maximal rate for the low $\tau$ and low $\phi$ portion of the MIPS state,
resulting in the emergence of a critical cluster regime.

At low $\phi$ there is no MIPS state because the collision frequency between
particles drops to values that are too low to sustain clusters. The clusters
are the most stable at the onset of MIPS and become less stable as $\phi$
increases.
Vacancies can nucleate anywhere inside an active solid and are not limited
to surface sites or favorable locations such as grain boundaries.
Given a crystalline patch of dense phase, there is some rate $R$
per unit surface area at which vacancies can nucleate per unit time.
As a
result, the total number $N_v$ of vacancies that nucleate in a given time
period is proportional to the total surface area $A$ of the cluster.
The fraction $C_L$ of particles in the largest
cluster remains nearly constant as a function of $\phi$ within the MIPS state,
and the dense phase is a close-packed crystal of particles so its
density does not vary with $\phi$.  
Thus $A$ must increase with increasing $\phi$
when $C_L$ remains constant since there are more particles
in the system.
The total number of vacancy nucleation attempts therefore increases
with increasing $\phi$. The nucleation success rate is determined by the
value of $\tau$ as described above.
The largest stable clusters appear when a low number of
nucleation attempts is combined with a low nucleation success rate,
and this situation occurs for intermediate
$\tau$ at densities that are just above the transition
into the MIPS state. For higher
values of $\tau$
or $\phi$, a larger number of vacancies successfully nucleate inside the active
crystal, where they can cluster into voids and destabilize the crystal
as a function of time, resulting in the emergence of the active crystal state.

\section{Summary}

We have analyzed motility-induced phase separation
in a two-dimensional active disk  
system using noise analysis and a machine learning principal component
analysis (PCA).
We show that 
PCA can capture time-dependent features
such as cluster rearrangements and breakups better
than standard structural measures such as the fraction of particles in
the largest cluster.
We also compute the power spectra of the time fluctuations in the
average speed of the particles
and find that in the active fluid, the noise is white and the noise
power is low,
but in the MIPS state, the noise has a $1/f^{\alpha}$ signature with
$\alpha = 1.6$,
similar to what is found in depinning systems or systems in a critical state.
We find that the PCA principal component and
the noise power can be combined
along with cluster size to characterize
different dynamic states within MIPS.
For intermediate persistence lengths and densities, we find that
the principal component and noise power
are large and there is a reentrant feature in the cluster size.
At
large densities and high persistence length the principal component is
low and the noise power is immediate, while in the
active fluid phase, all of the quantities are low.
From these measures, we propose that the MIPS state can be
divided into an active crystal  phase, a critical MIPS phase,
and a jammed phase, in addition to the fluid phase that occurs outside
of the MIPS regime.

  \acknowledgments
  This work was supported by the US Department of Energy through
  the Los Alamos National Laboratory.  Los Alamos National Laboratory is
  operated by Triad National Security, LLC, for the National Nuclear Security
  Administration of the U. S. Department of Energy (Contract No. 892333218NCA000001).
  Pacific University alumni Adrian Martin and Shannon Gallagher
  performed 
  related simulations and engaged in useful discussions.
  Their work was supported by
  the M. J. Murdock Charitable Trust.
  Computational resources include those from the 
  Center for Research Computing at the University of Notre Dame.
  
\bibliography{mybib}

\end{document}